\documentclass[aps,prb,twocolumn,superscriptaddress,showpacs]{revtex4}
\usepackage{graphicx}
\usepackage{dcolumn}
\usepackage{bm}
\usepackage{amssymb}
\usepackage{color}
\newcommand{\yco}{Y$_3$Co}
\newcommand{\rco}{$R_3$Co}

\begin{document}

\title{Temperature-driven Phase Transformation in Y$_3$Co: Neutron Scattering and First-principles Studies}

\author{A.~Podlesnyak}
\thanks{Corresponding author. Electronic address: podlesnyakaa@ornl.gov}
\author{G.~Ehlers}
\author{H.~Cao}
\author{M.~Matsuda}
\affiliation{Quantum Condensed Matter Division, Oak Ridge National Laboratory,
Oak Ridge, Tennessee 37831, USA}
\author{M.~Frontzek}
\author{O.~Zaharko}
\affiliation{Laboratory for Neutron Scattering, Paul Scherrer Institut, CH-5232 Villigen PSI, Switzerland}
\author{V.A.~Kazantsev}
\affiliation{Institute for Metal Physics RAS, 620041 Ekaterinburg, Russia}
\author{A.F.~Gubkin}
\author{N.V.~Baranov}
\affiliation{Institute for Metal Physics RAS, 620041 Ekaterinburg, Russia}
\affiliation{Institute of Natural Sciences, Ural Federal University, 620083 Ekaterinburg, Russia}
\date{\today}

\begin{abstract}
Contrary to previous studies that identified the ground state crystal structure of the entire \rco\ series ($R$ is a rare earth) as orthorhombic $Pnma$, we show that \yco\ undergoes a structural phase transition at $T_t \simeq 160$~K.
Single crystal neutron diffraction data reveal that at $T_t$ the trigonal prisms formed by a cobalt atom and its six nearest-neighbor yttrium atoms experience distortions accompanied by notable changes of the Y-Co distances.
The formation of the low-temperature phase is accompanied by a pronounced lattice distortion and anomalies seen in heat capacity and resistivity measurements.
Density functional theory calculations reveal a dynamical instability of the $Pnma$ structure of \yco.
In particular, a transversal acoustic phonon mode along the $(00\xi)$ direction has imaginary frequencies at $\xi < 1/4$.
Employing inelastic neutron scattering measurements we find a strong damping of the $(00\xi)$ phonon mode below a critical temperature $T_t$.
The observed structural transformation causes the reduction of dimensionality  of electronic bands and decreases the electronic density of states at the Fermi level that identifies \yco\ as a system with the charge density wave instability.
\end{abstract}


\maketitle

\section{INTRODUCTION}

The effects of a crystal structure transformation due to subtle atomic displacements have attracted much attention because they can lead to drastic changes in the electronic and magnetic properties of solids.
For example, in a number of multiferroic compounds the electric dipole moments typically develop from very small spontaneous deformations in the crystal structure.\cite{Kimura}
The cuprate high-temperature superconductors where superconductivity emerges in the proximity of, or in competition with, a charge-density wave (CDW) provide another example of materials where subtle atomic distortions play an essential role.\cite{Chang}
Currently most of the research on the subject is focused on strongly correlated oxides, although it is obvious that such phenomena could occur in intermetallic compounds as well.
Charge and spin density waves introduce a new periodicity in intermetallics, thereby breaking the lattice periodicity which defines their band structure.
In particular, the direct evidence of CDW formation was found in ternary rare-earth transition-metal silicides $R_5$Ir$_4$Si$_{10}$.\cite{Becker,Smaalen}
The results of the x-ray diffraction confirmed the existence of both incommensurate and commensurate CDW states in these compounds.
The formation of the CDW driven by the nesting of the Fermi surface and hosted within Te layers was detected in tritellurides $R$Te$_3$.\cite{Brouet,Sacchetti}
The reduction of the size of the Fermi surface at the transition has been studied extensively.\cite{Fawcett}
Anomalous in the electrical resistivity of nonmagnetic compounds Y$_2$CoIn$_8$, Y$_2$CoGa$_8$ and Dy$_2$CoGa$_8$ were also proposed to arise due to a CDW-induced anisotropic energy gap at the Fermi surface.\cite{Devang}
Moreover, high interest in tiny atomic displacements was reinforced recently when a purely electronic picture of CDW was stringently tested.\cite{Johannes,Zhu}
The coordinated action of electronic and ionic subsystems, i.e. the electron-phonon correlations was found to be responsible for the CDW instability in prototypical CDW materials CeTe$_3$ and MSe$_2$, M=Nb, Ta, Ti.

Binary intermetallic systems containing rare earth metals, such as \rco\ ($R$ is rare earth or yttrium) have been studied for more than three decades for their diversity of interesting properties.
They usually exhibit complicated magnetic behavior, depending on the rare-earth, including metamagnetic transitions,\cite{Primavesi,Gubkin} giant magnetoresistance,\cite{Baranov15} a substantial magnetocaloric effect\cite{Tripathy} and superconductivity.\cite{Geballe}
Co ions do not possess any ordered magnetic moment in \rco.\cite{Gubkin}

\yco\ with a nonmagnetic rare-earth site often serves as a reference system in systematic studies over the \rco\ series, but is also an interesting system in its own right.
A pronounced anomaly in the temperature dependent resistivity around $T_t \simeq 160$~K was reported more than twenty years ago but its origin was never satisfactorily explained.\cite{Talik0}
The anomaly was discussed in terms of interactions of the localized cobalt magnetic moments that is in contradiction with the later results of magnetic measurements.\cite{Baranov15}
A small hump of the specific heat curve around 160~K was also observed but never discussed.\cite{Baranov15}
The crystal structure was repeatedly reported to be orthorhombic Fe$_3$C-type structure ($Pnma$ space group) for entire \rco\ series at all temperatures.\cite{Cromer,Buschow}
Thus, the origin of the observed phenomena remained unclear.

In this work, using elastic and inelastic neutron scattering (INS) and density functional theory (DFT) calculations, we show that \yco\ undergoes a structural phase transition at $T_t \simeq 160$~K.
The formation of this new phase is evident by a pronounced lattice distortion, and heat capacity and resistivity anomalies.
We suggest that the nature of the observed anomalies at $T_t$ may be attributed to the formation of low-dimensional electronic bands with decreased density of states (DOS) at the Fermi level $\mathcal{N}(E_F)$ below the transition temperature.

\section{EXPERIMENTAL}

\yco\ samples were prepared by arc melting of the elements in a helium atmosphere. Several single crystals with largest dimensions of approximately $4 \times 4 \times 5$ mm$^3$ were grown by remelting the ingots at temperatures just above the peritectic point in a resistance furnace with a high temperature gradient, followed by annealing at 600$^{\circ}$C for 3 days.
Neutron single crystal diffraction experiments were performed at the HB-3A four circle diffractometer\cite{HB3A} at the High Flux Isotope Reactor (HFIR) and at the Cold Neutron Chopper Spectrometer (CNCS)\cite{CNCS} at the Spallation Neutron Source (SNS) both at Oak Ridge National Laboratory.
INS measurements were done using the thermal triple-axis spectrometer HB-1 at HFIR and at CNCS.
At HB-3A, measurements were done using a neutron wavelength of $\lambda=1.003$~\AA\ (Si-331 monochromator).
At HB-1 a neutron wavelength of 3.20~\AA\ was chosen with a PG-002 monochromator.
The incident wavelength at CNCS was $\lambda=1.80$~\AA.
The coefficients of linear thermal expansion $\alpha(T) = (1/L)(dL/dT)$ were measured using the dilatometer DL-1500 RHP/DL-1500-H (UCVAC/SINKU).
Heat capacity, resistivity and magnetization measurements were carried out using a Quantum Design PPMS.

\section{RESULTS AND DISCUSSION}

\begin{figure}[tb!]
\includegraphics[width=0.8\columnwidth]{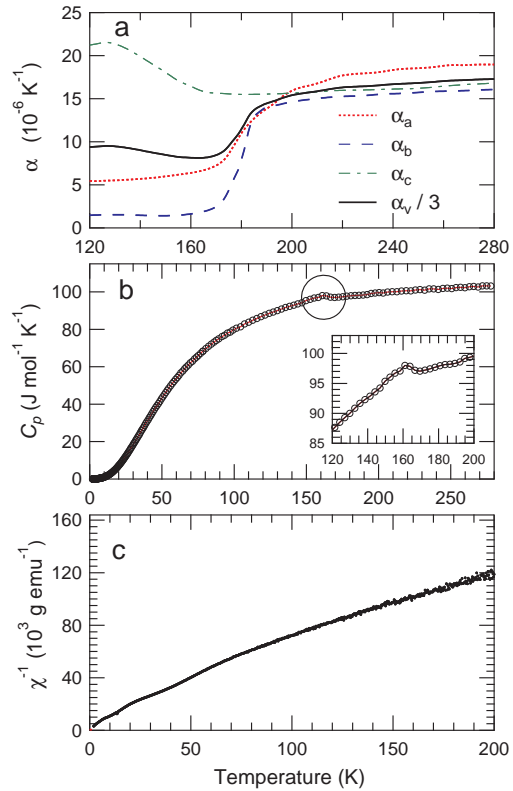}
\caption {(Color online) (a) The coefficients of linear thermal expansion $\alpha$ of Y$_3$Co as a function of temperature $T$ along the orthorhombic $a$, $b$ and $c$ axes. For comparison the volume expansion divided by a factor three ($\alpha_v/3$) is also shown. (b) Temperature dependence of the heat capacity of \yco\ with an anomaly at $T_t \simeq 160$~K. The insert shows data on an expanded scale around $T_t$. (c) Temperature dependence of the inverse magnetic susceptibility for a polycrystalline \yco.\label{cp}}
\end{figure}

Our single crystal exhibits an anomaly in its temperature dependence of the dc resistivity $\rho(T)$ similar to that published earlier in Refs.~\onlinecite{Talik0, Baranov15}.
For temperatures above 160 K, the resistivity in all three high-symmetry directions increases monotonically with temperature.
In the range $20-160$~K the resistivity curves, especially measured along the $c$ direction, exhibit a broad hump (not shown here, see Fig.~9 in Ref.~\onlinecite{Talik0}).
High-resolution measurements of the thermal expansion along the orthorhombic $a$, $b$ and $c$ axes reveal more pronounced anomalies.
A steplike decrease of the $\alpha_a$ and $\alpha_b$ coefficients  occurs around $T_t$ upon cooling whereas $\alpha_c$ slightly increases, see Fig.~\ref{cp}a.
As one can see in Fig.~\ref{cp}b a small but noticeable peak in the heat capacity curve is also visible at $T_t$.
Additional heat capacity measurements show that the anomaly has no magnetic field dependence.
Also, the inverse magnetic susceptibility curve does not show any peculiarities around 160~K (Fig.~\ref{cp}c).
Thus, a magnetic origin of the observed phenomena seems unlikely.
Rather, these observations suggest the presence of a temperature-driven structural phase transition in \yco\ at $T_t \simeq 160$~K.
Concerning the magnetic properties of \yco, the determination of basic mean-field parameters such as paramagnetic moments and the Curie-Weiss temperature is difficult (if not meaningless) because \yco\ does not follow a Curie-Weiss law over any long temperature interval.
This question was discussed in past, see for example Ref.~\onlinecite{Talik0}.

\begin{table*}[htb!]
\caption{\label{tab:1} Details of neutron diffraction data collection and crystal structure refinement.}
\begin{ruledtabular}
\begin{tabular}{lccccc}
&180 K&\multicolumn{4}{l}{120 K}\\
&$Pnma$&$Pnma$&$P2_12_12_1$&$P112_1$&$P1$\\
\hline
N of refined parameters& 11&11&17&32&62\\
$I>3\sigma(I)$/all&&&&&\\
Reflections collected&323/577&424/664&&&\\
Reflections unique&323/577&334/567&414/652&423/662&424/664\\
Reflections rejected&0/6&90/97&10/12&1/2&0/0\\
\hline
$I>3\sigma(I)$&&&&&\\
$R_1$/$wR_2$,\%&4.76/5.26&5.62/6.17&6.65/16.98&6.93/15.05&7.2/8.96\\
GooF&1.00&1.17&3.68&2.96&1.83\\
\end{tabular}
\end{ruledtabular}
\end{table*}

\begin{figure}[tb!]
\includegraphics[width=0.85\columnwidth]{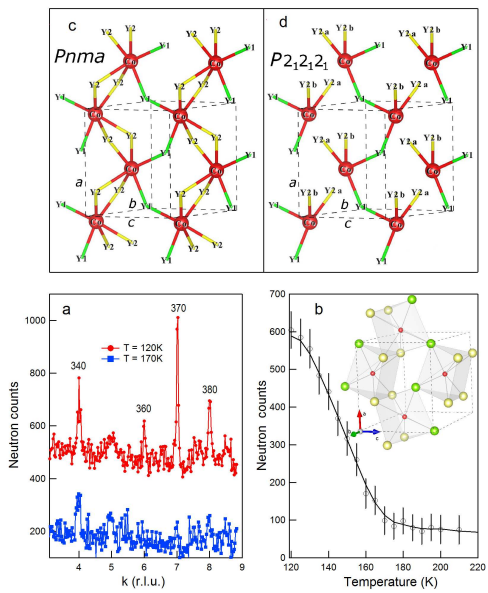}
\caption {(Color online) (a) A scan across $k$ of the $3k0$ reciprocal layer at $T=170$~K and 120~K. The 120~K pattern is shifted for clarity. (b) Temperature dependence of the integrated intensity of the (076) Bragg reflection which is symmetry-forbidden in $Pnma$ space group. Insert:  $Pnma$ crystal structure of \yco\ above $T_t$. Green, yellow, and red spheres represent Y1, Y2, and Co ions, respectively. Only one of two layers of the trigonal prisms is shown for clarity. (c,d) Schematic representation of the main deformation of the \yco\ crystal structure upon $Pnma \rightarrow P 2_1 2_1 2_1$ phase transition. The Y1-Co bonds remain rather uniform across the transition, while the Y2-Co distances change. This breaks Co-Y2-Co layer-connected chains. \label{diffr}}
\end{figure}

First, we use elastic neutron scattering to explore structural transformations.
Using the CNCS time of flight spectrometer we acquire multiple datasets for different orientations of the crystal and combine them in software to fully map out the scattering function $S(\mathbf{Q},E)$ in a large $\mathbf{Q}$ region.
A sector of 100 degree rotation with 1$^{\circ}$ step using $(hk0)$ scattering plane was recorded in total.
A number of weak reflections were observed in the diffraction pattern measured at $T=120$~K which are absent at $T=170$~K, as demonstrated in Fig.~\ref{diffr}a.
The data were obtained by integration in the $h$ direction from 2.95 to 3.05 reciprocal lattice units (r.l.u.) and in  energy transfer from -0.1 to 0.1~meV.
These reflections violate the extinction rules for the Bragg reflections in $Pnma$ space group.
Fig.~\ref{diffr}b shows a typical temperature dependence of the integrated peak intensity of one of the strongest forbidden reflections.

The observation of a few very weak, symmetry-forbidden reflections stimulated further investigations using the four circle neutron diffractometer HB-3A.
A total of 577 and 664 reflections were collected at $T=180$~K and 120~K, respectively.
A structure refinement of the diffraction data was carried out with the Jana2000 program,\cite{jana} the details are presented in Table~\ref{tab:1}.
Our criteria in choice of the model were i) fulfillment of symmetry conditions of a space group and ii) quality of the refinement defined by the conventional parameters:
$R_1 = (\Sigma||F_o|-|F_c||) / \Sigma|F_o|$,
$wR_2 =\sqrt{ \Sigma[w(F_o^2-F_c^2)^2] / \Sigma[w(F_o^2)^2] }$
and GooF $=\sqrt{ \Sigma[w(F_o^2-F_c^2)^2 ] / (n-p) }$.
Here $F_o$ and $F_c$ are observed and calculated structure factors, $w$ is the weight $1/\sigma(F_o)^2$, $n$ is the number of reflections and $p$ is the total number of refined parameters.

The orthorhombic $Pnma$ crystal structure with lattice parameters $a=7.024(5)$~\AA, $b=9.429(6)$~\AA\ and $c=6.358(5)$~\AA\ was confirmed for the high temperature phase.
From 577 reflections measured at 180 K only 6 weak observations ($I<3\sigma(I)$) are rejected as systematically extinct.
Also the R-values of refinement are good.
In this structure twelve yttrium ions per unit cell occupy two nonequivalent positions, 4$c$ (Y1) and 8$d$ (Y2).
Four Co ions are located at the 4$c$ position within the trigonal prisms formed by yttrium ions.
The prisms are interconnected through edge- and corner-sharing to form layers as shown in the insert of Fig.~\ref{diffr}b.

From 664 reflections measured at 120 K 90 significant observations ($I>3\sigma(I)$) are rejected as systematically extinct.
Therefore, though the R-values of refinement in the $Pnma$ space group are acceptable, the resulting parameters correspond only to the average structure.
The fine details of the structural distortion at $T_t$ can be extracted only when the symmetry conditions satisfying the new reflections are taken into account.
All reflections can be indexed only in  the lowest symmetry group  $P1$.
However, the number of relevant peaks and their intensities are at present insufficient for a full structure determination.
The atomic coordinates and isotropic displacement parameters obtained for different space groups are summarized in Table~\ref{tab:2}.

\begin{table*}[!ht]
\caption{\label{tab:2}Atomic coordinates and isotropic displacement parameters for Y$_3$Co.}
\begin{ruledtabular}
\begin{tabular}{llllllllll}
& $x$ & $y$ & $z$ & $U_{iso}$ [\AA$^2$]& &$x$ & $y$ & $z$ & $U_{iso}$ [\AA$^2$]\\
&\multicolumn{4}{c}{$T=180$~K $Pnma$}&&\multicolumn{4}{c}{$T=120$~K $Pnma$} \\
&\multicolumn{4}{c}{$a$=7.024(5) $b$=9.429(6) $c$=6.358(5) \AA}&&\multicolumn{4}{c}{$a$=7.021(5) $b$=9.428(6) $c$=6.354(5) \AA}\\
\hline
Co  & 0.388(1) & 1/4 & 0.960(1) & 0.018(1)& Co &0.388(1) & 1/4 & 0.960(1) & 0.019(2) \\
Y1  & 0.0425(3) & 1/4 & 0.1334(4) & 0.0144(5)&Y1  & 0.0429(4) & 1/4 & 0.1339(4) & 0.0126(5) \\
Y2  & 0.1759(2) & 0.0651(2) & 0.6760(2) & 0.0158(4)& Y2  & 0.1763(3) &0.0657(2) & 0.6761(3) & 0.0150(4)\\
\hline
&\multicolumn{4}{c}{$T=120$~K $P2_12_12_1$}&\multicolumn{4}{c}{$T=120$~K $P111$} \\
Co  & ~0.389(3) & ~0.250 & ~0.960(3) & 0.020(4) & Co1a  & ~0.3928 & ~0.5007 & ~0.9601 & 0.012(3)\\
Y1  & ~0.04218(8) & ~0.2496(1) & ~0.1334(9) & 0.014(1) & Co1b  & ~0.884(2) & -0.509(1) & -0.462(4) & 0.014(3)\\
Y2a  & ~0.1806(5) & ~0.0685(3) & ~0.6676(6) & 0.0163(8) & Co1c  & -0.378(4) & ~0.007(2) &-0.955(5) & 0.023(4)\\
Y2b  & -0.1718(5) & -0.0645(3) & -0.6842(8) & 0.0163(8) & Co1d  & -0.892(4) & ~0.002(2) & ~0.465(4) & 0.019(3)\\
&&&&& Y1a  & ~0.047(4) & ~0.472(2) & ~0.110(5) & 0.020(1)\\
&\multicolumn{4}{c}{$T=120$~K $P112_1$} & Y1b  & ~0.552(4) & -0.520(2) & ~0.349(5) & 0.015(1)\\
&&&& & Y1c  & -0.030(3) & ~0.982(2) & -0.158(4) & 0.0107(9)\\
Co1a  & 0.393(2) & ~0.5007(9) & ~0.960 & 0.020(3) & Y1d  & -0.531(3) & -0.019(2) & -0.388(4) & 0.0106(9)\\
Co1b  & 0.881(2) & -0.5043(9) & -0.462(6) & 0.018(3) & Y2a  & ~0.188(3) & ~0.306(2) & ~0.669(4) & 0.014(1)\\
Y1a  & 0.0416(8) & ~0.4979(3)  & ~0.140(8) & 0.015(1) & Y2b  & ~0.685(3) & -0.322(2) & -0.194(5) & 0.025(2)\\
Y1b  & 0.5391(8) & -0.5040(3) & ~0.373(8) & 0.014(1) & Y2c  & -0.161(3) & ~0.805(2) & -0.695(4) & 0.011(1)\\
Y2a  & 0.1728(8) & ~0.3148(4)  & ~0.686(8) & 0.0169(9) & Y2d  & -0.661(3) & ~0.174(2) & ~0.179(5) & 0.024(1) \\
Y2b  & 0.6721(8) & -0.3136(4) & -0.182(8) & 0.0159(9) & Y3a  & ~0.694(3) & ~0.305(2) & ~0.815(4) & 0.009(1)\\
Y3a  & 0.6800(8) & ~0.3166(4)  & ~0.832(8) & 0.0140(9)&Y3b  & ~0.201(3) &-0.328(2) & -0.354(4) & 0.027(2)\\
Y3b  & 0.1789(8) & -0.3194(4) & -0.332(8) & 0.0176(9)&Y3c  & -0.662(3) & ~0.805(2) & -0.850(4) & 0.0073(9)\\
&&&&&Y3d  & -0.165(3) & ~0.175(2) & ~0.318(5) & 0.017(1) \\
\end{tabular}
\end{ruledtabular}
\end{table*}

It turns out that the measured integrated intensities at $T=120$~K could best described within the space group $P 2_1 2_1 2_1$.
Here the R-values are higher than for the $Pnma$ and $P1$ models, but advantage of the $P2_12_12_1$ model is that almost all reflections are taken into account and changes of the structure from the $Pnma$ model can be easily followed.
Fig.~\ref{diffr}c-d present the Y-Co bonds within one layer.
The Co and Y1 atoms have almost the same positions in the $Pnma$ and $P2_12_12_1$ models, whereas the Y2 position is split into the Y2a and Y2b sites in $P2_12_12_1$ and these atoms are shifted maximal in $c$ (by $\approx$0.05 \AA) and minimal in $b$ (by $\approx$0.02 \AA) from the average position.
The main consequence of the structural transition is spanning of the Co-Y2a, Co-Y2b interatomic distances, their variation reaches almost 5\%, see Table~\ref{tab:3} .

\begin{table}[!ht]
\caption{\label{tab:3}Selected bond lengths [\AA] for Y$_3$Co above and below the transition temperature $T_t$.}
\begin{ruledtabular}
\begin{tabular}{lllll}
&$T=180$~K, $Pnma$&$T=120$~K, $P2_12_12_1$\\
\hline
Co - Y1  & 2.666(8) &  2.66(1) \\
Co - Y1  & 2.802(8) &  2.82(2)\\
Co - Y2a  & 2.806(6) & 2.84(1)\\
Co - Y2a  & 2.921(6) & 2.88(1)\\
Co - Y2b  & & 2.79(1) \\
Co - Y2b  & & 2.93(1) \\
\end{tabular}
\end{ruledtabular}
\end{table}

A possible origin of the temperature-driven structural phase transition observed in \yco\ will now be discussed.
To understand the phase stability of the \yco\ high-temperature structure, it is essential to know the lattice dynamics of the $Pnma$ phase.
It is well known that at high temperatures a number of elements and intermetallic compounds adopt energetically unstable crystal structures.
For example, the transition metals of groups III and IV are stable in a bcc phase at high temperatures and transform to hcp upon lowering the temperature.\cite{Petry}
They exhibit one or several dynamically unstable phonon modes which are characterized by imaginary harmonic frequencies.\cite{Souvatzis}
These phonon modes are dynamical precursors of the ionic displacements and cause the transition when the temperature is lowered.
To check for such a scenario we have studied the dynamical stability of the $Pnma$ structure of \yco\ at $T=0$~K.
First-principles calculations were performed using the Vienna \emph{ab initio} simulation package (VASP)\cite{vasp} based on projector augmented wave (PAW) pseudopotentials \cite{Kresse} within the generalized gradient approximation (GGA) as parameterized by Perdew, Burke, and Ernzerhof (PBE).\cite{Perdew}
After careful convergence tests, a plane wave cutoff energy of 400~eV and an $11 \times 11 \times 11$ $k$-point Monkhorst-Pack mesh\cite{Monkhorst} for the 16-atom unit cell of \yco\ were found to be sufficient to converge the total energy to better than 1~meV/atom.
The resulting optimized lattice constants ($a = 6.9801$~\AA, $b = 9.4894$~\AA, $c = 6.2810$~\AA) are in good agreement with our experimental values, within $\sim{0.6}-1$\%.
Phonon spectra were computed by means of the supercell approach with the finite displacement method\cite{Parlinski} realized in Phonopy package.\cite{phonopy}
We used the 128 atomic $2 \times 2 \times2 $ supercell created from the optimized primitive unit cell.
The computational cost of these calculations becomes unreasonable high, due to the low symmetry, if a larger cell is considered.

\begin{figure}[tb!]
\includegraphics[width=0.95\columnwidth]{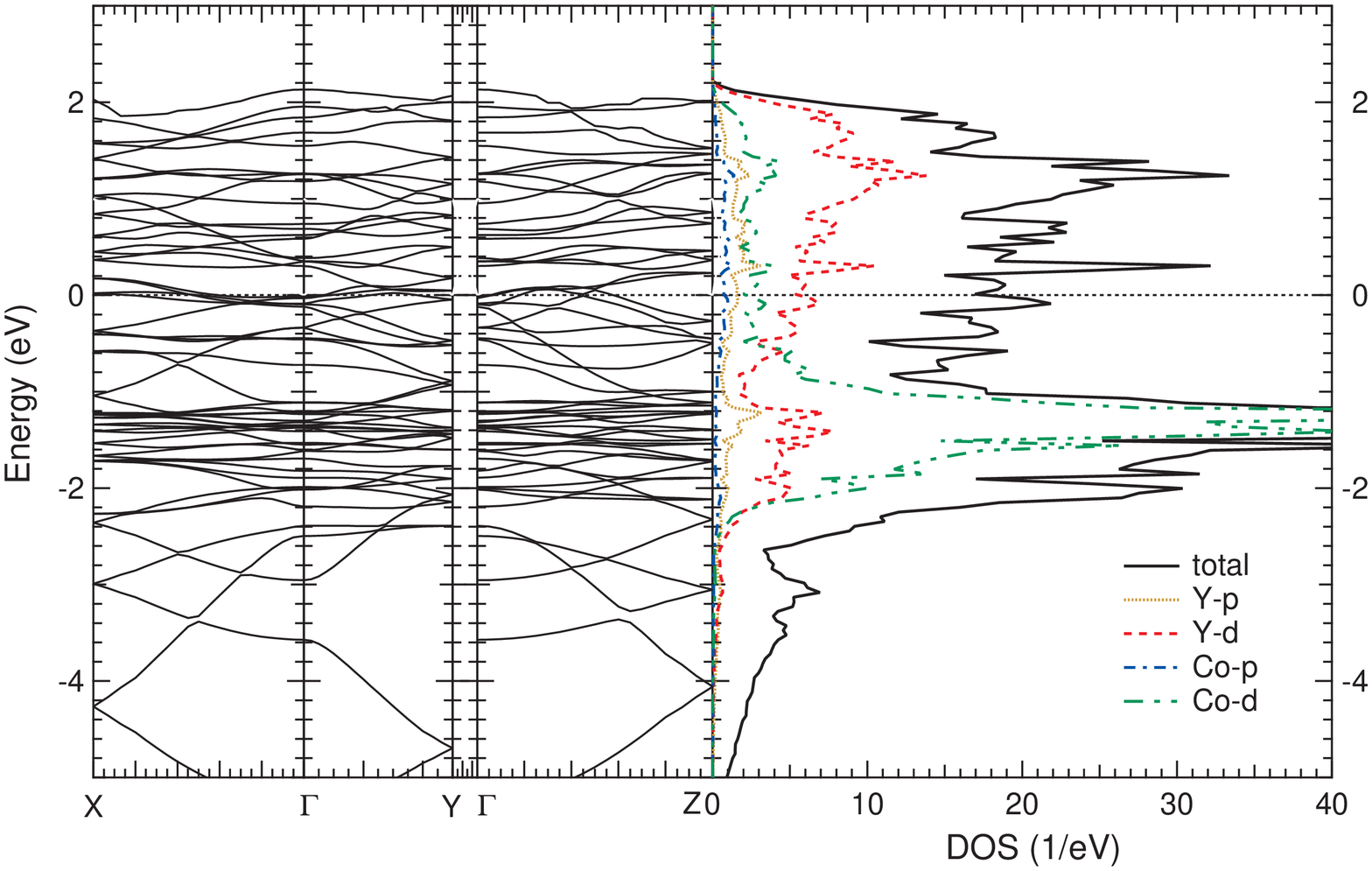}
\caption {(Color online)  Left: Energy bands along high symmetry directions of the first Brillouin zone of \yco. The horizontal line corresponds to the Fermi energy shifted to zero reference level. The positions of $\Gamma$, $X$, $Y$ and $Z$ correspond to (0,0,0), (1/2,0,0), (0,1/2,0) and (0,0,1/2), respectively. Right: partial and total DOS of Y and Co in \yco.\label{edos}}
\end{figure}

The PAW-GGA-DFT based energy bands along high-symmetry directions of the Brillouin zone and DOS of \yco\ are shown in Fig.~\ref{edos}.
As evident from the Fig.~\ref{edos}, energy bands around $E_F$ arise mainly due to Y 4d and Co 3d states, with a dominance of Y 4d electrons.
It turns out that the Fermi level situates close to a DOS minimum between Co 3d and Y 4d bands, that means that the second derivative of the density of state is positive.
The combination of first and second derivatives 3[$1/\mathcal{N}$ d$^2\mathcal{N}/$d$E^2$]$^2_{E=E_F}$ - [$1/\mathcal{N}$ d$\mathcal{N} /$d$E$]$_{E=E_F}$ could explain the tendency of temperature dependence of resistivity $\rho(T)$ to saturate, see for instance Ref.~\onlinecite{Jones}.

\begin{figure}[tb!]
\includegraphics[width=0.95\columnwidth]{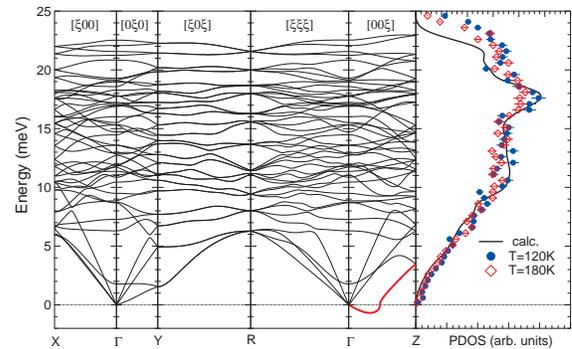}
\caption {(Color online) Left: calculated phonon dispersion curves of \yco\ in $Pnma$ space group along various high-symmetry directions in the Brillouin zone (with the negative axis convention used for imaginary frequencies, see text). Right: calculated (solid line) and experimental (symbols) phonon density of states determined from the INS spectra of \yco\ measured at CNCS at temperatures 120~K and 180~K.\label{vasp}}
\end{figure}

The calculated phonon dispersion relations with 48 phonon branches along various high-symmetry directions in the Brillouin zone and the corresponding phonon density of states (PDOS) are shown in Fig.~\ref{vasp}.
The calculated PDOS reproduces well the frequencies of the peaks and shoulders of the experimental data.
We do not observe a statistically significant difference between PDOS taken at temperatures above and below the phase transition.
However, as one can see in Fig.~\ref{vasp} the resulting phonon band structure of the $Pnma$ phase exhibits a clear instability, evidenced by the appearance of the imaginary phonon frequencies (represented in Fig.~\ref{vasp} as negative values) of the lowest acoustic branch in the $[00\xi]$ direction with $\xi<1/4$.
Analysis of the eigenvectors of the unstable acoustic phonon mode shows that the polarization vector is polarized along the $[0\xi0]$ direction. This suggests that the Y2a and Y2b ions of the Co-Y2-Co layer-connected chains displace in opposite directions, towards the structure stabilized at low temperature.

\begin{figure}[tb!]
\includegraphics[width=0.86\columnwidth]{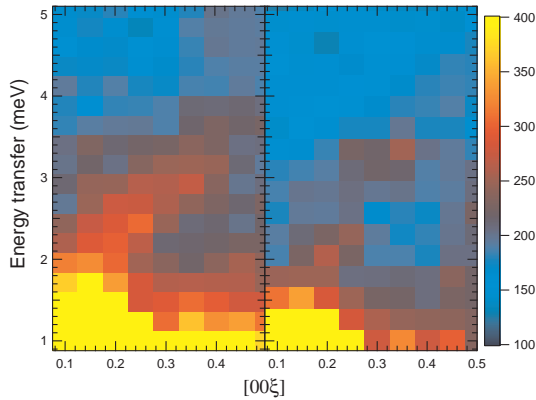}
\caption {(Color online) Contour map from constant $E$ scans of Y$_3$Co measured at $T=180$~K (left panel) and $T=120$~K (right panel). \label{phon}}
\end{figure}

\begin{figure}[t!]
\includegraphics[width=0.85\columnwidth]{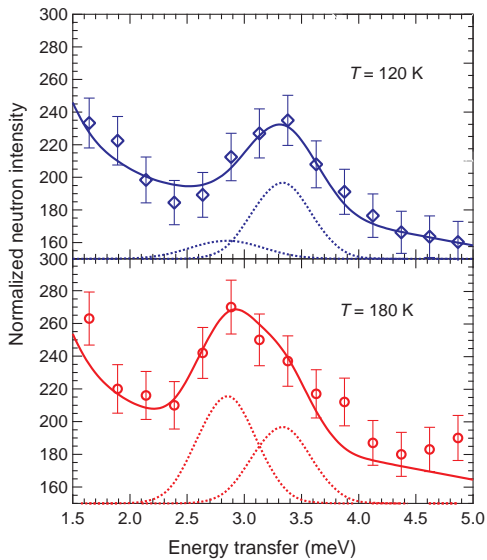}
\caption {(Color online) Constant $\mathbf{q}$ scans at $\mathbf{q} = [0, 0, 0.35]$ corrected with Bose temperature factor, above and below transition. Solid lines are the results of fit with two resolution limited Gaussians and sloped background. Dash lines are the individual components of the spectra.\label{cut}}
\end{figure}

We further focus on the predicted dynamically unstable $[00\xi]$ acoustic mode at temperatures above and below the phase transition.
We constructed intensity contour maps based on a series of constant-energy scans obtained at the triple-axis spectrometer HB-1, see Fig.~\ref{phon}.
A strong reduction of the $(00\xi)$ phonon mode intensity below a critical temperature $T_t$ is clearly seen.
Fig.~\ref{cut} shows constant $\mathbf{q}$ scans at $\mathbf{q} = [0, 0, 0.35]$ for temperatures above and below phase transition.
In order to compare the phonon intensities measured at different temperatures the data were corrected by the Bose population factor $[n(\omega)+1]$, $n(\omega)=[\exp\{\hbar \omega / k_{\mathrm{B}} T\} -1]^{-1}$.
According to the calculated phonon dispersions the INS spectrum in $[00\xi]$ direction at energy transfer $E<5$~meV comprises two acoustic modes including dynamically unstable one.
These modes cannot be resolved in our measurements due to finite resolution but at $T=180$~K can be modeled by two resolution limited Gaussians, see Fig.~\ref{cut}.
Lowering the temperature to 120~K we found that the lower Gaussian is almost vanished whereas upper mode is unaffected.
The peak broadening was not observed in our measurements.
This indicates that indeed the transversal acoustic (TA) low-energy $[00\xi]$ phonon branch is unstable and vanishes at temperatures below the phase transition.

Overall, we can argue that (i) the anomaly in heat capacity which happens at the same temperature $T_t \simeq 160$~K as an anomalous thermal expansion along $c$, (ii) the maximal shift of the Y2a and Y2b in $c$ direction in comparison with $a$ and $b$ and (iii) unstable low-energy $[00\xi]$ phonon branch, indicate of the lattice instability along $c$ direction.
Comparing our results with those from known CDW compounds with low-dimensional electronic bands we find qualitative similarities.
Most of the CDW systems have anomalies in the thermodynamic properties at their critical temperature $T_{CDW}$.
For instance, the transitions in a strong interchain coupled CDW systems $R_5$Ir$_4$Si$_{10}$  are characterized by a sharp jumps in the susceptibility and the resistivity as well as by narrow cusp in the specific heat.\cite{Becker,Smaalen}
We can therefore speculate that the observed structural distortions at $T_t$ in \yco\ also cause the reduction of the electronic bands dimensionality and a decrease of DOS at the Fermi level.

We should also note that our recent single crystal neutron diffraction measurements on the Tb$_3$Co intermetallic compound reveal a superstructure pattern similar that was found in \yco.\cite{Matthias}
We suggest that some other members of the \rco\ series (or even all of them) have ground state crystal symmetry lower than $Pnma$.

\section{CONCLUSIONS}

To summarize, using advanced neutron scattering techniques, we were able to analyze the temperature-driven structural transition in the \yco\ and to determine the basic distortions occurring at critical temperature $T_t$.
Using inelastic neutron scattering data and \emph{ab initio} calculations we have shown that at low temperatures the $Pnma$ orthorhombic structure has a dynamically unstable transversal acoustic phonon mode, which induces the ionic displacements.
We argue, the structure transformation causes the reduction of  dimensionality of electronic bands and decreases the DOS at the Fermi level that places this system among the CDW materials.
Thus, our results support the theoretical conclusion that CDW formation cannot be a purely electronic effect \cite{Johannes,Zhu}.
It generally relies on the lattice distortion as an essential element and requires a strong $q$-dependent electron-phonon coupling.

\begin{acknowledgments}
The research at Oak Ridge National Laboratory$'$s High Flux Isotope Reactor and Spallation Neutron Source was sponsored by the Scientific User Facilities Division, Office of Basic Energy Sciences, US Department of Energy (DOE).
This research used resources of the National Energy Research Scientific Computing Center, which is supported by the Office of Science of the U.S. Department of Energy under Contract No. DE-AC02-05CH11231.
This work was partly supported by the program of the Ural Branch of RAS (project No. 12-T-2-1012) and by the Ministry of Education and Science of the Russian Federation (Contract 14.518.11.7020).
\end{acknowledgments}

\bibliographystyle{apsrev}


\begin{thebibliography}{31}
\expandafter\ifx\csname natexlab\endcsname\relax\def\natexlab#1{#1}\fi
\expandafter\ifx\csname bibnamefont\endcsname\relax
  \def\bibnamefont#1{#1}\fi
\expandafter\ifx\csname bibfnamefont\endcsname\relax
  \def\bibfnamefont#1{#1}\fi
\expandafter\ifx\csname citenamefont\endcsname\relax
  \def\citenamefont#1{#1}\fi
\expandafter\ifx\csname url\endcsname\relax
  \def\url#1{\texttt{#1}}\fi
\expandafter\ifx\csname urlprefix\endcsname\relax\def\urlprefix{URL }\fi
\providecommand{\bibinfo}[2]{#2}
\providecommand{\eprint}[2][]{\url{#2}}

\bibitem[{\citenamefont{Kimura et~al.}(2003)\citenamefont{Kimura, Goto,
  Shintani, Ishizaka, Arima, and Tokura}}]{Kimura}
\bibinfo{author}{\bibfnamefont{T.}~\bibnamefont{Kimura}},
  \bibinfo{author}{\bibfnamefont{T.}~\bibnamefont{Goto}},
  \bibinfo{author}{\bibfnamefont{H.}~\bibnamefont{Shintani}},
  \bibinfo{author}{\bibfnamefont{K.}~\bibnamefont{Ishizaka}},
  \bibinfo{author}{\bibfnamefont{T.}~\bibnamefont{Arima}}, \bibnamefont{and}
  \bibinfo{author}{\bibfnamefont{Y.}~\bibnamefont{Tokura}},
  \bibinfo{journal}{Nature (London)} \textbf{\bibinfo{volume}{426}},
  \bibinfo{pages}{55} (\bibinfo{year}{2003}).

\bibitem[{\citenamefont{Chang et~al.}(2012)\citenamefont{Chang, Blackburn,
  Holmes, Christensen, Larsen, Mesot, Liang, Bonn, Hardy, A.Watenphul
  et~al.}}]{Chang}
\bibinfo{author}{\bibfnamefont{J.}~\bibnamefont{Chang}},
  \bibinfo{author}{\bibfnamefont{E.}~\bibnamefont{Blackburn}},
  \bibinfo{author}{\bibfnamefont{A.~T.} \bibnamefont{Holmes}},
  \bibinfo{author}{\bibfnamefont{N.~B.} \bibnamefont{Christensen}},
  \bibinfo{author}{\bibfnamefont{J.}~\bibnamefont{Larsen}},
  \bibinfo{author}{\bibfnamefont{J.}~\bibnamefont{Mesot}},
  \bibinfo{author}{\bibfnamefont{R.}~\bibnamefont{Liang}},
  \bibinfo{author}{\bibfnamefont{D.~A.} \bibnamefont{Bonn}},
  \bibinfo{author}{\bibfnamefont{W.~N.} \bibnamefont{Hardy}},
  \bibinfo{author}{\bibnamefont{A.Watenphul}}, \bibnamefont{et~al.},
  \bibinfo{journal}{Nature Physics} \textbf{\bibinfo{volume}{8}},
  \bibinfo{pages}{871} (\bibinfo{year}{2012}).

\bibitem[{\citenamefont{Becker et~al.}(1999)\citenamefont{Becker, Patil,
  Ramakrishnan, Menovsky, Nieuwenhuys, Mydosh, Kohgi, and Iwasa}}]{Becker}
\bibinfo{author}{\bibfnamefont{B.}~\bibnamefont{Becker}},
  \bibinfo{author}{\bibfnamefont{N.~G.} \bibnamefont{Patil}},
  \bibinfo{author}{\bibfnamefont{S.}~\bibnamefont{Ramakrishnan}},
  \bibinfo{author}{\bibfnamefont{A.~A.} \bibnamefont{Menovsky}},
  \bibinfo{author}{\bibfnamefont{G.~J.} \bibnamefont{Nieuwenhuys}},
  \bibinfo{author}{\bibfnamefont{J.~A.} \bibnamefont{Mydosh}},
  \bibinfo{author}{\bibfnamefont{M.}~\bibnamefont{Kohgi}}, \bibnamefont{and}
  \bibinfo{author}{\bibfnamefont{K.}~\bibnamefont{Iwasa}},
  \bibinfo{journal}{Phys. Rev. B} \textbf{\bibinfo{volume}{59}},
  \bibinfo{pages}{7266} (\bibinfo{year}{1999}).

\bibitem[{\citenamefont{van Smaalen et~al.}(2004)\citenamefont{van Smaalen,
  Shaz, Palatinus, Daniels, Galli, Nieuwenhuys, and Mydosh}}]{Smaalen}
\bibinfo{author}{\bibfnamefont{S.}~\bibnamefont{van Smaalen}},
  \bibinfo{author}{\bibfnamefont{M.}~\bibnamefont{Shaz}},
  \bibinfo{author}{\bibfnamefont{L.}~\bibnamefont{Palatinus}},
  \bibinfo{author}{\bibfnamefont{P.}~\bibnamefont{Daniels}},
  \bibinfo{author}{\bibfnamefont{F.}~\bibnamefont{Galli}},
  \bibinfo{author}{\bibfnamefont{G.~J.} \bibnamefont{Nieuwenhuys}},
  \bibnamefont{and} \bibinfo{author}{\bibfnamefont{J.~A.}
  \bibnamefont{Mydosh}}, \bibinfo{journal}{Phys. Rev. B}
  \textbf{\bibinfo{volume}{69}}, \bibinfo{pages}{014103}
  (\bibinfo{year}{2004}).

\bibitem[{\citenamefont{Brouet et~al.}(2004)\citenamefont{Brouet, Yang, Zhou,
  Hussain, Ru, Shin, Fisher, and Shen}}]{Brouet}
\bibinfo{author}{\bibfnamefont{V.}~\bibnamefont{Brouet}},
  \bibinfo{author}{\bibfnamefont{W.~L.} \bibnamefont{Yang}},
  \bibinfo{author}{\bibfnamefont{X.~J.} \bibnamefont{Zhou}},
  \bibinfo{author}{\bibfnamefont{Z.}~\bibnamefont{Hussain}},
  \bibinfo{author}{\bibfnamefont{N.}~\bibnamefont{Ru}},
  \bibinfo{author}{\bibfnamefont{K.~Y.} \bibnamefont{Shin}},
  \bibinfo{author}{\bibfnamefont{I.~R.} \bibnamefont{Fisher}},
  \bibnamefont{and} \bibinfo{author}{\bibfnamefont{Z.~X.} \bibnamefont{Shen}},
  \bibinfo{journal}{Phys. Rev. Lett.} \textbf{\bibinfo{volume}{93}},
  \bibinfo{pages}{126405} (\bibinfo{year}{2004}).

\bibitem[{\citenamefont{Sacchetti et~al.}(2009)\citenamefont{Sacchetti,
  Condron, Gvasaliya, Pfuner, Lavagnini, Baldini, Toney, Merlini, Hanfland,
  Mesot et~al.}}]{Sacchetti}
\bibinfo{author}{\bibfnamefont{A.}~\bibnamefont{Sacchetti}},
  \bibinfo{author}{\bibfnamefont{C.~L.} \bibnamefont{Condron}},
  \bibinfo{author}{\bibfnamefont{S.~N.} \bibnamefont{Gvasaliya}},
  \bibinfo{author}{\bibfnamefont{F.}~\bibnamefont{Pfuner}},
  \bibinfo{author}{\bibfnamefont{M.}~\bibnamefont{Lavagnini}},
  \bibinfo{author}{\bibfnamefont{M.}~\bibnamefont{Baldini}},
  \bibinfo{author}{\bibfnamefont{M.~F.} \bibnamefont{Toney}},
  \bibinfo{author}{\bibfnamefont{M.}~\bibnamefont{Merlini}},
  \bibinfo{author}{\bibfnamefont{M.}~\bibnamefont{Hanfland}},
  \bibinfo{author}{\bibfnamefont{J.}~\bibnamefont{Mesot}},
  \bibnamefont{et~al.}, \bibinfo{journal}{Phys. Rev. B}
  \textbf{\bibinfo{volume}{79}}, \bibinfo{pages}{201101}
  (\bibinfo{year}{2009}).

\bibitem[{\citenamefont{Fawcett}(1988)}]{Fawcett}
\bibinfo{author}{\bibfnamefont{E.}~\bibnamefont{Fawcett}},
  \bibinfo{journal}{Rev. Mod. Phys.} \textbf{\bibinfo{volume}{60}},
  \bibinfo{pages}{209} (\bibinfo{year}{1988}).

\bibitem[{\citenamefont{Devang et~al.}(2009)\citenamefont{Devang, Nigam, Dhar,
  and Thamizhavel}}]{Devang}
\bibinfo{author}{\bibfnamefont{A.~J.} \bibnamefont{Devang}},
  \bibinfo{author}{\bibfnamefont{A.~K.} \bibnamefont{Nigam}},
  \bibinfo{author}{\bibfnamefont{S.~K.} \bibnamefont{Dhar}}, \bibnamefont{and}
  \bibinfo{author}{\bibfnamefont{A.}~\bibnamefont{Thamizhavel}},
  \bibinfo{journal}{Phys. Rev. B} \textbf{\bibinfo{volume}{80}},
  \bibinfo{pages}{054414} (\bibinfo{year}{2009}).

\bibitem[{\citenamefont{Johannes and Mazin}(2008)}]{Johannes}
\bibinfo{author}{\bibfnamefont{M.~D.} \bibnamefont{Johannes}} \bibnamefont{and}
  \bibinfo{author}{\bibfnamefont{I.~I.} \bibnamefont{Mazin}},
  \bibinfo{journal}{Phys. Rev. B} \textbf{\bibinfo{volume}{77}},
  \bibinfo{pages}{165135} (\bibinfo{year}{2008}).

\bibitem[{\citenamefont{Zhu et~al.}(2012)\citenamefont{Zhu, Cheng, and
  Schwingenschl\"ogl}}]{Zhu}
\bibinfo{author}{\bibfnamefont{Z.}~\bibnamefont{Zhu}},
  \bibinfo{author}{\bibfnamefont{Y.}~\bibnamefont{Cheng}}, \bibnamefont{and}
  \bibinfo{author}{\bibfnamefont{U.}~\bibnamefont{Schwingenschl\"ogl}},
  \bibinfo{journal}{Phys. Rev. B} \textbf{\bibinfo{volume}{85}},
  \bibinfo{pages}{245133} (\bibinfo{year}{2012}).

\bibitem[{\citenamefont{Primavesi and Taylor}(1972)}]{Primavesi}
\bibinfo{author}{\bibfnamefont{G.}~\bibnamefont{Primavesi}} \bibnamefont{and}
  \bibinfo{author}{\bibfnamefont{K.~N.~R.} \bibnamefont{Taylor}},
  \bibinfo{journal}{J. Phys. F: Met. Phys.} \textbf{\bibinfo{volume}{2}},
  \bibinfo{pages}{761} (\bibinfo{year}{1972}).

\bibitem[{\citenamefont{Gubkin et~al.}(2010)\citenamefont{Gubkin, Podlesnyak,
  and Baranov}}]{Gubkin}
\bibinfo{author}{\bibfnamefont{A.~F.} \bibnamefont{Gubkin}},
  \bibinfo{author}{\bibfnamefont{A.}~\bibnamefont{Podlesnyak}},
  \bibnamefont{and} \bibinfo{author}{\bibfnamefont{N.~V.}
  \bibnamefont{Baranov}}, \bibinfo{journal}{Phys. Rev. B}
  \textbf{\bibinfo{volume}{82}}, \bibinfo{pages}{012403}
  (\bibinfo{year}{2010}).

\bibitem[{\citenamefont{Baranov et~al.}(2001)\citenamefont{Baranov, Yermakov,
  Markin, Possokhov, Michor, Weingartner, and Hilscher}}]{Baranov15}
\bibinfo{author}{\bibfnamefont{N.~V.} \bibnamefont{Baranov}},
  \bibinfo{author}{\bibfnamefont{A.~A.} \bibnamefont{Yermakov}},
  \bibinfo{author}{\bibfnamefont{P.~E.} \bibnamefont{Markin}},
  \bibinfo{author}{\bibfnamefont{U.~M.} \bibnamefont{Possokhov}},
  \bibinfo{author}{\bibfnamefont{H.}~\bibnamefont{Michor}},
  \bibinfo{author}{\bibfnamefont{B.}~\bibnamefont{Weingartner}},
  \bibnamefont{and} \bibinfo{author}{\bibfnamefont{G.}~\bibnamefont{Hilscher}},
  \bibinfo{journal}{J. Alloys and Compounds} \textbf{\bibinfo{volume}{329}},
  \bibinfo{pages}{22} (\bibinfo{year}{2001}).

\bibitem[{\citenamefont{Tripathy et~al.}(2006)\citenamefont{Tripathy, Suresh,
  and Nigam}}]{Tripathy}
\bibinfo{author}{\bibfnamefont{S.~K.} \bibnamefont{Tripathy}},
  \bibinfo{author}{\bibfnamefont{K.~G.} \bibnamefont{Suresh}},
  \bibnamefont{and} \bibinfo{author}{\bibfnamefont{A.~K.} \bibnamefont{Nigam}},
  \bibinfo{journal}{J. Magn. Magn. Mater.} \textbf{\bibinfo{volume}{306}},
  \bibinfo{pages}{24} (\bibinfo{year}{2006}).

\bibitem[{\citenamefont{Geballe et~al.}(1965)\citenamefont{Geballe, Matthias,
  Compton, Correnzwit, Hull, Jr., and Longinotti}}]{Geballe}
\bibinfo{author}{\bibfnamefont{T.~H.} \bibnamefont{Geballe}},
  \bibinfo{author}{\bibfnamefont{B.~T.} \bibnamefont{Matthias}},
  \bibinfo{author}{\bibfnamefont{V.~B.} \bibnamefont{Compton}},
  \bibinfo{author}{\bibfnamefont{E.}~\bibnamefont{Correnzwit}},
  \bibinfo{author}{\bibfnamefont{G.~W.} \bibnamefont{Hull}},
  \bibinfo{author}{\bibnamefont{Jr.}}, \bibnamefont{and}
  \bibinfo{author}{\bibfnamefont{L.~D.} \bibnamefont{Longinotti}},
  \bibinfo{journal}{Phys. Rev.} \textbf{\bibinfo{volume}{137}},
  \bibinfo{pages}{A119} (\bibinfo{year}{1965}).

\bibitem[{\citenamefont{Talik et~al.}(1988)\citenamefont{Talik, Szade, Heimann,
  Winiarska, Winiarski, and Chelkowski}}]{Talik0}
\bibinfo{author}{\bibfnamefont{E.}~\bibnamefont{Talik}},
  \bibinfo{author}{\bibfnamefont{J.}~\bibnamefont{Szade}},
  \bibinfo{author}{\bibfnamefont{J.}~\bibnamefont{Heimann}},
  \bibinfo{author}{\bibfnamefont{A.}~\bibnamefont{Winiarska}},
  \bibinfo{author}{\bibfnamefont{A.}~\bibnamefont{Winiarski}},
  \bibnamefont{and}
  \bibinfo{author}{\bibfnamefont{A.}~\bibnamefont{Chelkowski}},
  \bibinfo{journal}{J. Less-Common Metals} \textbf{\bibinfo{volume}{138}},
  \bibinfo{pages}{129} (\bibinfo{year}{1988}).

\bibitem[{\citenamefont{Cromer and Larson}(1961)}]{Cromer}
\bibinfo{author}{\bibfnamefont{D.~T.} \bibnamefont{Cromer}} \bibnamefont{and}
  \bibinfo{author}{\bibfnamefont{A.~C.} \bibnamefont{Larson}},
  \bibinfo{journal}{Acta Crystallogr.} \textbf{\bibinfo{volume}{14}},
  \bibinfo{pages}{1226} (\bibinfo{year}{1961}).

\bibitem[{\citenamefont{Buschow and van~der Goot}(1969)}]{Buschow}
\bibinfo{author}{\bibfnamefont{K.~H.~J.} \bibnamefont{Buschow}}
  \bibnamefont{and} \bibinfo{author}{\bibfnamefont{A.~S.} \bibnamefont{van~der
  Goot}}, \bibinfo{journal}{J. Less-Common Met.} \textbf{\bibinfo{volume}{18}},
  \bibinfo{pages}{309} (\bibinfo{year}{1969}).

\bibitem[{\citenamefont{Chakoumakos et~al.}(2011)\citenamefont{Chakoumakos,
  Cao, Ye, Stoica, Popovici, Sundaram, Zhou, Hicks, Lynn, and Riedel}}]{HB3A}
\bibinfo{author}{\bibfnamefont{B.~C.} \bibnamefont{Chakoumakos}},
  \bibinfo{author}{\bibfnamefont{H.}~\bibnamefont{Cao}},
  \bibinfo{author}{\bibfnamefont{F.}~\bibnamefont{Ye}},
  \bibinfo{author}{\bibfnamefont{A.~D.} \bibnamefont{Stoica}},
  \bibinfo{author}{\bibfnamefont{M.}~\bibnamefont{Popovici}},
  \bibinfo{author}{\bibfnamefont{M.}~\bibnamefont{Sundaram}},
  \bibinfo{author}{\bibfnamefont{W.}~\bibnamefont{Zhou}},
  \bibinfo{author}{\bibfnamefont{J.~S.} \bibnamefont{Hicks}},
  \bibinfo{author}{\bibfnamefont{G.~W.} \bibnamefont{Lynn}}, \bibnamefont{and}
  \bibinfo{author}{\bibfnamefont{R.~A.} \bibnamefont{Riedel}},
  \bibinfo{journal}{J. Appl. Cryst.} \textbf{\bibinfo{volume}{44}},
  \bibinfo{pages}{655} (\bibinfo{year}{2011}).

\bibitem[{\citenamefont{Ehlers et~al.}(2011)\citenamefont{Ehlers, Podlesnyak,
  Niedziela, Iverson, and Sokol}}]{CNCS}
\bibinfo{author}{\bibfnamefont{G.}~\bibnamefont{Ehlers}},
  \bibinfo{author}{\bibfnamefont{A.}~\bibnamefont{Podlesnyak}},
  \bibinfo{author}{\bibfnamefont{J.~L.} \bibnamefont{Niedziela}},
  \bibinfo{author}{\bibfnamefont{E.~B.} \bibnamefont{Iverson}},
  \bibnamefont{and} \bibinfo{author}{\bibfnamefont{P.~E.} \bibnamefont{Sokol}},
  \bibinfo{journal}{Rev. Sci. Instrum.} \textbf{\bibinfo{volume}{82}},
  \bibinfo{pages}{085108} (\bibinfo{year}{2011}).

\bibitem[{jan()}]{jana}
\bibinfo{howpublished}{V. Petricek, M. Dusek and L. Palatinus (2000): Jana2000.
  The crystallographic computing system. Institute of Physics, Praha, Czech
  Republic}.

\bibitem[{\citenamefont{Petry et~al.}(1991)\citenamefont{Petry, Heiming,
  Trampenau, Alba, Herzig, Schober, and Vogl}}]{Petry}
\bibinfo{author}{\bibfnamefont{W.}~\bibnamefont{Petry}},
  \bibinfo{author}{\bibfnamefont{A.}~\bibnamefont{Heiming}},
  \bibinfo{author}{\bibfnamefont{J.}~\bibnamefont{Trampenau}},
  \bibinfo{author}{\bibfnamefont{M.}~\bibnamefont{Alba}},
  \bibinfo{author}{\bibfnamefont{C.}~\bibnamefont{Herzig}},
  \bibinfo{author}{\bibfnamefont{H.~R.} \bibnamefont{Schober}},
  \bibnamefont{and} \bibinfo{author}{\bibfnamefont{G.}~\bibnamefont{Vogl}},
  \bibinfo{journal}{Phys. Rev. B} \textbf{\bibinfo{volume}{43}},
  \bibinfo{pages}{10933} (\bibinfo{year}{1991}).

\bibitem[{\citenamefont{Souvatzis et~al.}(2008)\citenamefont{Souvatzis,
  Eriksson, Katsnelson, and Rudin}}]{Souvatzis}
\bibinfo{author}{\bibfnamefont{P.}~\bibnamefont{Souvatzis}},
  \bibinfo{author}{\bibfnamefont{O.}~\bibnamefont{Eriksson}},
  \bibinfo{author}{\bibfnamefont{M.~I.} \bibnamefont{Katsnelson}},
  \bibnamefont{and} \bibinfo{author}{\bibfnamefont{S.~P.} \bibnamefont{Rudin}},
  \bibinfo{journal}{Phys. Rev. Lett.} \textbf{\bibinfo{volume}{100}},
  \bibinfo{pages}{095901} (\bibinfo{year}{2008}).

\bibitem[{\citenamefont{Kresse and Furthmuller}(1996)}]{vasp}
\bibinfo{author}{\bibfnamefont{G.}~\bibnamefont{Kresse}} \bibnamefont{and}
  \bibinfo{author}{\bibfnamefont{J.}~\bibnamefont{Furthmuller}},
  \bibinfo{journal}{Phys. Rev. B} \textbf{\bibinfo{volume}{54}},
  \bibinfo{pages}{11169} (\bibinfo{year}{1996}).

\bibitem[{\citenamefont{Kresse and Joubert}(1999)}]{Kresse}
\bibinfo{author}{\bibfnamefont{G.}~\bibnamefont{Kresse}} \bibnamefont{and}
  \bibinfo{author}{\bibfnamefont{J.}~\bibnamefont{Joubert}},
  \bibinfo{journal}{Phys. Rev. B} \textbf{\bibinfo{volume}{59}},
  \bibinfo{pages}{1758} (\bibinfo{year}{1999}).

\bibitem[{\citenamefont{Perdew et~al.}(1996)\citenamefont{Perdew, Burke, and
  Ernzerhof}}]{Perdew}
\bibinfo{author}{\bibfnamefont{J.~P.} \bibnamefont{Perdew}},
  \bibinfo{author}{\bibfnamefont{K.}~\bibnamefont{Burke}}, \bibnamefont{and}
  \bibinfo{author}{\bibfnamefont{M.}~\bibnamefont{Ernzerhof}},
  \bibinfo{journal}{Phys. Rev. Lett.} \textbf{\bibinfo{volume}{77}},
  \bibinfo{pages}{3865} (\bibinfo{year}{1996}).

\bibitem[{\citenamefont{Monkhorst and Pack}(1972)}]{Monkhorst}
\bibinfo{author}{\bibfnamefont{H.~J.} \bibnamefont{Monkhorst}}
  \bibnamefont{and} \bibinfo{author}{\bibfnamefont{J.~D.} \bibnamefont{Pack}},
  \bibinfo{journal}{Phys. Rev. B} \textbf{\bibinfo{volume}{13}},
  \bibinfo{pages}{5188} (\bibinfo{year}{1972}).

\bibitem[{\citenamefont{Parlinski et~al.}(1997)\citenamefont{Parlinski, Li, and
  Kawazoe}}]{Parlinski}
\bibinfo{author}{\bibfnamefont{K.}~\bibnamefont{Parlinski}},
  \bibinfo{author}{\bibfnamefont{Z.~Q.} \bibnamefont{Li}}, \bibnamefont{and}
  \bibinfo{author}{\bibfnamefont{Y.}~\bibnamefont{Kawazoe}},
  \bibinfo{journal}{Phys. Rev. Lett.} \textbf{\bibinfo{volume}{78}},
  \bibinfo{pages}{4063} (\bibinfo{year}{1997}).

\bibitem[{\citenamefont{Togo et~al.}(2008)\citenamefont{Togo, Oba, and
  Tanaka}}]{phonopy}
\bibinfo{author}{\bibfnamefont{A.}~\bibnamefont{Togo}},
  \bibinfo{author}{\bibfnamefont{F.}~\bibnamefont{Oba}}, \bibnamefont{and}
  \bibinfo{author}{\bibfnamefont{I.}~\bibnamefont{Tanaka}},
  \bibinfo{journal}{Phys. Rev. B} \textbf{\bibinfo{volume}{78}},
  \bibinfo{pages}{134106} (\bibinfo{year}{2008}).

\bibitem[{\citenamefont{Jones}(1956)}]{Jones}
\bibinfo{author}{\bibfnamefont{H.}~\bibnamefont{Jones}},
  \emph{\bibinfo{title}{Hdb. Phys.}} (\bibinfo{publisher}{Berlin: Springer
  Verlag}, \bibinfo{year}{1956}), vol.~\bibinfo{volume}{2}, p.
  \bibinfo{pages}{266}.

\bibitem[{Mat()}]{Matthias}
\bibinfo{howpublished}{M. Frontzek \textit{et al}. (to be published).}

\end{thebibliography}

\end{document}